\documentclass[showpacs,preprintnumbers, twocolumn,
amsmath,amssymb,APSl,prd,nofootinbib,superscriptaddress]{revtex4-2}

\usepackage{amsmath}
\usepackage{amssymb}
\usepackage{amsfonts}
\usepackage{graphicx,bm}
\usepackage{dcolumn}
\usepackage[colorlinks=true]{hyperref}
\usepackage{color,amsxtra}
\usepackage{epsf}
\usepackage{enumerate}
\usepackage{hhline}
\usepackage{array}
\usepackage{tabularx}
\usepackage{subfigure}
\usepackage{fancyhdr}
\usepackage{mathrsfs}

\newcommand{\be}{\begin{equation}}
\newcommand{\ee}{\end{equation}}
\newcommand{\bea}{\begin{eqnarray}}
\newcommand{\eea}{\end{eqnarray}}
\newcommand{\beaa}{\begin{eqnarray*}}
\newcommand{\eeaa}{\end{eqnarray*}}

\newcommand{\nn}{\nonumber \\}

\def\be{\begin{equation}}
\def\ee{\end{equation}}
\def\bea{\begin{eqnarray}}
\def\eea{\end{eqnarray}}

\begin{document}

\title{Modified gravity/Dynamical Dark Energy vs $\Lambda$CDM: is the game over?}

\author{Sergei D. Odintsov}
\email{odintsov@ice.csic.es} \affiliation{Institut de Ci\`{e}ncies de l'Espai,
ICE/CSIC-IEEC, Campus UAB, Carrer de Can Magrans s/n, 08193 Bellaterra (Barcelona),
Spain}
 \affiliation{Instituci\'o Catalana de Recerca i Estudis Avan\c{c}ats (ICREA),
Passeig Luis Companys, 23, 08010 Barcelona, Spain}
 \author{Diego~S\'aez-Chill\'on~G\'omez}
\email{diego.saez@uva.es}
\affiliation{Department of Theoretical, Atomic and Optical
Physics, Campus Miguel Delibes, \\ University of Valladolid UVA, Paseo Bel\'en, 7, 47011
Valladolid, Spain}
\affiliation{Department of Physics, Universidade Federal do Cear\'a (UFC), Campus do Pici, Fortaleza - CE, C.P. 6030, 60455-760 - Brazil}
\author{German~S.~Sharov}
 \email{sharov.gs@tversu.ru}
 \affiliation{Tver state university, Sadovyj per. 35, 170002 Tver, Russia}
 \affiliation{International Laboratory for Theoretical Cosmology,
Tomsk State University of Control Systems and Radioelectronics (TUSUR), 634050 Tomsk,
Russia}

%

%
\begin{abstract}

Over the last decades, tests on the standard model of cosmology, the so-called
$\Lambda$CDM model, have been widely analysed and compared with many different models
for describing dark energy. Modified gravities have played an important role in this
sense as an alternative to $\Lambda$CDM model. Previous observational data has been
always favouring $\Lambda$CDM model in comparison to any other model. While
statistically speaking, alternative models have shown tehir power, fitting in some cases the
observational data slightly better than $\Lambda$CDM, the significance and goodness of
the fits were not significantly relevant to exclude the standard model of
cosmology. In this paper, a generalisation of exponential $F(R)$ gravity is considered and compared with $\Lambda$CDM model by using the latest observational data. Also some well-known model independent parameterisations for the equation of state (EoS) of dark energy are explored. These scenarios are confronted with the renewed observational
data involving the Pantheon plus datasets of supernovae  type Ia,  the
Hubble parameter estimations, data from the cosmic microwave background and
baryon acoustic oscillations, where the latter includes the data provided by Dark Energy Spectroscopic Instrument collaboration.
 Results of this analysis suggest that standard exponential $F(R)$ models provide much better fits than $\Lambda$CDM model, which is excluded at 4$\sigma$. Moreover, the parameterisations of the equation of state suggest a non-constant EoS parameter for dark energy, where $\Lambda$CDM model is also excluded at  4$\sigma$.



%

\end{abstract}
\maketitle
%
%

\section{Introduction}

Modern cosmology aims to approach an exhaustive explanation of available observational data. To do so, many different approaches are followed in the literature. The Cosmological Standard model assumes General Relativity (GR) to be valid but at the price of assuming additional unknown components, such as dark matter, dark
energy. Other well-known ways to explain the observations lies on  considering different types of modifications of
GR \cite{reviews1}. From time to time cosmologists have to improve the corresponding theories in order to achieve to describe new release of observational data.

Recent measurements of baryon acoustic oscillations (BAO) from the  Dark Energy
Spectroscopic Instrument (DESI) from Data Release 1 point to corrections in the predictions of some known cosmological models \cite{DESI:2024}. The DESI Data Release 1
includes BAO data from clustering of galaxies, quasars and the Lyman-$\alpha$ forest in the redshift range $0.1<z<4.16$. For some popular cosmological scenarios
these new data in combination with the Pantheon+ compilation of type Ia supernovae
(SN Ia) data \cite{PantheonP:2022} and other observational datasets can improve
the estimations of the free model parameters, particularly the equation of state for dark
energy, the Hubble constant as well as other parameters \cite{DESI:2024,CosmolDESI:2024}. Moreover, new data, particularly late-time data, might help to infer about the solution to the Hubble constante tension \cite{DiValentino:2021izs}, which seems to be eluded when one just considers early-time data \cite{Vagnozzi:2023nrq}.

In this paper, the new BAO and SN Ia observations are considered in the framework of modified gravity. In particular, the so-called  $F(R)$ gravity, an extension of GR with a Lagrangian that depends non-trivially on the Ricci scalar $R$ is analysed. $F(R)$ gravity models have been widely analysed in the literature because they can successfully describe both the inflationary era and the late-time
epoch with accelerating expansion of the universe \cite{reviews1,reviews2}.
In addition, some observational limitations that include local constraints, as in the Solar system, together with a reasonable limit to recover the usual cosmological evolution along radiation and matter dominated epoch have been analysed and some viable models that accomplish some particular constraints have been considered in the literature \cite{Hu:2007nk,Nojiri:2007cq,Linder2009,Appleby:2007vb}, which can also describe well
the SN Ia, BAO, CMB observational data related to the late-time cosmological evolution \cite{Cruz-Dombriz:2015,Ravi:2023nsn,Oliveros:2023ewl}.
\\

One of the most interesting models of this type of viable $F(R)$ gravities is a class of
exponential gravities \cite{Linder2009,nostriexp}. These models contain a term that
consists on a negative exponential of the Ricci scalar $R$, such that it mimics
$\Lambda$CDM model asymptotically (at early times), when the exponential becomes
negligible. Such scenario generally occurs in two cases: (a) as far as the Ricci scalar
is large enough at the early epoch and/or (b) for a large value of the parameter that
modulates the exponential. Indeed, when the Ricci scalar is below a particular scale,
the exponential term plays an important role, and the resulting cosmological evolution
can essentially differ from that predicted by the $\Lambda$CDM model.  As shown
previously in the literature, this type of exponential gravity  can successfully
describe observational data  in comparison with other models
\cite{YangLeeLG2010,OdintsovSGS:2017,OdintsovSGStens:2021}. Moreover, exponential
gravities may reproduce the whole cosmological evolution, unifying early-time inflation
and late-time acceleration. For this purpose an action should include an inflationary
term (for example, based on $R^2$ inflation) that act effectively at early times but is
suppressed or vanishes at the end of inflation \cite{nostriexp,OdintsovSGS:2017}. Some
modifications of exponential gravity with extra logarithmic terms motivated by quantum
gravity and with an axion scalar field have been also studied and tested versus
observational data in some previous papers
\cite{OdintsovOS:2017log,OdintsovSGSlog:2019,OdintsovSGS_Axi:2023}. Some other generalisations of 
exponential $F(R)$ gravity have been considered, where a term of the type $e^{f(R)}$ is included in the  action \cite{Oliveros:2023ewl,Granda:2020,Oikonomou:2020}.


In this paper, the standard exponential gravity model proposed in Ref.~\cite{Linder2009,nostriexp} is considered together with a generalisation of such a model. Both are tested with observational data and compared to $\Lambda$CDM model.  Results of these tests
 strongly favour exponential $F(R)$ gravity models in comparison
to $\Lambda$CDM model, even when considering the fact that the number of free parameters
is larger for the $F(R)$ gravity cases. In this sense, $\Lambda$CDM model is excluded at more than 3$\sigma$ for both exponential gravity models, reaching 4$\sigma$ in the case of standard exponential gravity. Moreover, in order to check and support such results, some well-known model agnostic parameterisations for the dark energy equation of state (EoS) are considered, namely:  $w$CDM and
Chevallier-Polarski-Linder (CPL) models \cite{CPL}, where the former assumes a constant EoS parameter and the latter a dynamical one. Results also favour in this case a dynamical EoS for dark energy versus a constant one. Indeed, $w$CDM provides better fits than $\Lambda$CDM model, which lies within the $2\sigma$ confidence region i.e. with no statistically relevance, whereas CPL models fit much better the data and excludes $\Lambda$CDM at 4$\sigma$. These analysis confirm previous conclusions by some recent papers
 \cite{DESI:2024,Chudaykin:2024,Park:2024brx,Notari:2024} where the recent DESI BAO data points to an evidence for modified gravity or
dynamical dark energy. In addition, despite other older analysis also favoured the possibility of modified gravity \cite{Nunes:2016qyp}, the evidence now turns out much stronger. Such an evidence is also supported by other different analysis where CMB anisotropy power spectra is studied \cite{DiValentino:2020hov}. Moreover, for long time the possibility that the EoS parameter deviates from $-1$ has been pointed out, also by some recent analysis \cite{Escamilla:2023oce}, such that all these results support the idea that $\Lambda$CDM model might be ruled out definitely.



The paper is organised as follows: in section \ref{Dynamics}, the dynamical equations
for the exponential $F(R)$ gravity are obtained. In section \ref{Data},  SN Ia, BAO,
$H(z)$ and CMB observational data are briefly described. The models are confronted with
the observations and results are analysed in section \ref{Results}. In the next section
we compare our  $F(R)$ gravity tests with  $w$CDM and CPL scenarios. Finally,
conclusions are provided in section \ref{conclusions}.

\section{Dynamics in $F(R)$ gravity}
\label{Dynamics}

The so-called $F(R)$ gravity models are described by the following gravitational action:
\cite{Linder2009,nostriexp,YangLeeLG2010,OdintsovSGS:2017}
\begin{equation}
  S =\int d^4x \sqrt{-g}\bigg[ \frac{F(R)}{2\kappa^2}  + {\cal L}_{m} \bigg]\, ,
 \label{Act1}\end{equation}
where $ {\cal L}_{m}$ is the matter Lagrangian, $\kappa^2=8\pi G$ with  $G$ being the
Newtonian gravitational constant. Here, we are assuming a flat
Friedmann-Lema\^itre-Robertson-Walker metric:
\be
 ds^2 = -dt^2 + a^2(t)\,\delta_{ij}dx^idx^j\ .
 \label{FLRW} \ee
Then, by varying the action \eqref{Act1}, FLRW equations are obtained \cite{OdintsovSGS:2017,OdintsovSGSlog:2019}:
 \begin{eqnarray}
 3 H^2F_R&=&\frac{RF_R-F}{2}-3H\dot{F}_R+\kappa^2\rho\, , \nn
-2\dot{H}F_R&=&\ddot{F}_R-H\dot{F}_R +\kappa^2(\rho+p)\, .\label{eqn2}
 \end{eqnarray}
Here $F_R\equiv \frac{dF(R)}{dR}$, $H=\dot a/a$ is the Hubble parameter and the dot denotes
derivatives with respect to the cosmic time $t$, whereas $\rho$ and $p$ are the energy density and
the pressure for all the species (including dark matter and all the species), respectively. The
continuity equation for the energy-momentum tensor takes the usual form:
 \begin{equation}
 \dot\rho=-3H(\rho+p)\ .
  \label{cont}\end{equation}

For the purposes of this paper, we might use the relation $R=6\dot H + 12H^2$ to rewrite the FLRW
equations (\ref{eqn2}) as a dynamical system as follows:
\cite{OdintsovSGS:2017,OdintsovSGS_Axi:2023}:
 \begin{eqnarray}
\frac{dH}{d\log a}&=&\frac{R}{6H}-2H\ , \label{dynam1} \\
\frac{dR}{d\log
a}&=&\frac1{F_{RR}}\bigg(\frac{\kappa^2\rho}{3H^2}-F_R+\frac{RF_R-F}{6H^2}\bigg)\ .
 \label{dynam2}
 \end{eqnarray}

Here, we are considering a particular model of $F(R)$  gravity, given by:
 \begin{equation}
 F(R)=R -2\Lambda  \big(1-e^{-\beta{\cal R}^\alpha}\big)\ ,
\qquad  {\cal R}=\frac{R}{2\Lambda}\ ,
   \label{FR2}
\end{equation}
 where $\beta$ is a dimensionless parameter and
${\cal R}$ is the normalized Ricci scalar. As shown in some previous literature, an additional term $F_\mathrm{inf}$ might be considered in the action \eqref{FR2}, which would correspond to the dominated term during inflation \cite{OdintsovSGS:2017}. However, such inflationary contribution is considered tp become negligible at late-times, such that is not considered in this paper. Note also that by setting $\alpha=1$ in (\ref{FR2}), one recovers the well-known exponential $F(R)$ gravity that has been considered in previous literature
\cite{Linder2009,nostriexp,OdintsovSGS:2017}. In addition, for $\alpha>0$ the model (\ref{FR2}) reduces to the usual $\Lambda$CDM model for the limits
$\beta\to+\infty$ and/or $R\to+\infty$, where the latter turns out at early times. In this sense, the analysis of this paper considers observational data (see Sect.~\ref{Data} for more details) for late times, since the earliest data comes from
the Cosmic Microwave Background (CMB), which is located at redshifts $z\simeq1100$. Other observational data, as SN Ia, CC and BAO, are located at much less redshifts $z\le2.4$. The epoch with redshifts $z<10^4$ corresponds to ${\cal R}<10^{13}$ when any
inflationary term $F_\mathrm{inf}$ becomes negligible, since the normalised Ricci scalar at
the end of inflation is many orders of magnitude larger, ${\cal R}_0\sim 10^{85}$
\cite{OdintsovSGS_LnAx:2024}.
During the matter domination epoch at $z<10^4$, pressureless matter contains baryons
and dark matter: $\rho_m= \rho_b+\rho_{dm}$. Then, cosmological evolution for the matter and radiation energy
densities are obtained by the continuity equation, which yield:
(\ref{cont}):
\begin{equation}
 \rho= \rho_m^0a^{-3}+ \rho_r^0a^{-4},
\label{rho}\end{equation}
where $\rho_m^0$ and $\rho_r^0$ are the energy densities at the
present time $t_0$, where $a(t_0)=1$. In order to reduce the number of free parameters, the radiation-matter ratio is fixed by Planck data \cite{OdintsovSGS_Axi:2023,OdintsovSGS_LnAx:2024,Planck13} as:
   \begin{equation}
X_r=\frac{\rho_r^0}{\rho_m^0}=2.9656\cdot10^{-4}\ .
 \label{Xrm} \end{equation}
As pointed above, for the limit $R\to+\infty$ or more precisely,  for $\beta{\cal R}^\alpha\gg1$, the model (\ref{FR2}) becomes close to  $\Lambda$CDM model, such that one can assume with no loss of generality that the Hubble parameter and the Ricci scalar would be close to the ones given for the
$\Lambda$CDM model asymptotically, which are given by:
\cite{OdintsovSGS:2017,OdintsovSGS_Axi:2023,OdintsovSGS_LnAx:2024}:
 \begin{equation}
 \frac{H^2}{H^{*2}_0}=\Omega_m^{*} \big(a^{-3}+ X_r a^{-4}\big)+\Omega_\Lambda^{*}\,,\quad
 \frac{R}{2\Lambda}=2+\frac{\Omega_m^{*}}{2\Omega_\Lambda^{*}}a^{-3}\ .
  \label{asymLCDM}\end{equation}
Here $H^{*}_0$, $\Omega_m^{*}$ and $\Omega_\Lambda^{*}$  are the Hubble constant, matter/$\Lambda$ cosmological parameters
for the $\Lambda$CDM model that mimics the exponential $F(R)$ model at
asymptotically at early times. Nevertheless, in general, the values for the cosmological parameters will differ for the $F(R)$ gravity model
(\ref{FR2}), since even under the assumption that the $F(R)$ model matches
(\ref{asymLCDM}) at high redshifts $10^3\le z\le10^5$, the late-time evolution will
deviate from the $\Lambda$CDM scenario. In this sense, the parameters for both models are related as follows \cite{OdintsovSGS_Axi:2023}:
\bea
 \Omega_m^0H_0^2&=&\Omega_m^{*}(H^{*}_0)^2=\frac{\kappa^2}3\rho_m(t_0)\ , \nn
 \qquad  \Omega_\Lambda H_0^2&=&\Omega_\Lambda^{*}(H^{*}_0)^2=\frac{\Lambda}3\ .
  \label{H0Omm}\eea
Then, by using $H^*_0$, the Hubble parameter can be normalised as $E=\frac{H}{H_0^{*}}$. Whereas  the dynamical equations (\ref{dynam1}) and (\ref{dynam2}) can be expressed in terms of the
dimensionless variables $E(a)$ and ${\cal R}(a)$, which lead to:
\begin{widetext}
\begin{eqnarray}
\frac{dE}{dx}&=&\Omega_\Lambda^{*}\frac{{\cal R}}{E}-2E,\qquad x=\log a\ , \label{eqE} \\   
\frac{d{\cal R}}{dx}&=&\frac{\big[\Omega_m^{*}(a^{-3}+ X_r
a^{-4})+\Omega_\Lambda^{*}\big(1-(1+\alpha\beta{\cal R}^{\alpha})\,e^{-\beta{\cal
R}^\alpha}\big)\big]\big/E^2-1+\alpha\beta{\cal R}^{\alpha-1} e^{-\beta{\cal R}^\alpha}}
 {\alpha\beta(\alpha\beta{\cal R}^{\alpha}+1-\alpha)\,{\cal R}^{\alpha-2} e^{-\beta{\cal R}^\alpha}}\,.
  \label{eqR}
  \end{eqnarray}
\end{widetext}

By following some previous analysis (see for instance \cite{OdintsovSGS:2017,OdintsovSGS_Axi:2023}) the system of equations (\ref{eqE})-(\ref{eqR}) can be solved numerically by starting at some appropriate
initial redshift where the  $\Lambda$CDM asymptotical conditions (\ref{asymLCDM}) hold, which are then imposed as initial conditions for the equations. This means that at the initial point
 the factor $\epsilon=e^{-\beta {\cal R_{ini}}^\alpha}$ should be much smaller than unity. For the calculations of this paper, $\epsilon\sim
10^{-9}$ is assumed, which can be used to obtain the integration starting point through Eq.~(\ref{asymLCDM}), which gives:
\be
x_\mathrm{ini}=-\frac13\log\bigg\{\frac{2\Omega_\Lambda^*}{\Omega_m^*}\bigg[\bigg(\frac{\log\epsilon^{-1}}{\beta}\bigg)^{1/\alpha}-2\bigg]\bigg\}\ .
 \ee
Thus, the solutions $E(a)$, ${\cal R}(a)$ can be obtained for the particular $F(R)$
model that is considered here, while the Hubble parameter $H(a)$ or $H(z)$ is obtained via Eqs.~(\ref{H0Omm}). Remind that $z=a^{-1}-1$ and $a(t_0)=1$. Hence, the cosmological evolution for this $F(R)$ model can be compared with observational data, which is described in the next section.

\section{Observational data}
\label{Data}

In order to fit the model (\ref{FR2}) with observational data, Supernovae Ia (SNe Ia), baryon acoustic oscillations (BAO), estimations of the
Hubble parameter $H(z)$ or Cosmic Chronometers (CC)  and parameters from the cosmic
microwave background radiation (CMB) are considered.

In this paper, the Pantheon+ sample database \cite{PantheonP:2022} is used, which
provides $N_{\mathrm{SN}}=1701$ datapoints that contains information of the distance moduli $\mu_i^\mathrm{obs}$
at redshifts $z_i$ from 1550 spectroscopically SNe Ia. Then, the $\chi^2$ function is computed:
\bea
\chi^2_{\mathrm{SN}}(\theta_1,\dots)&=&\min\limits_{H_0} \sum_{i,j=1}^{N_\mathrm{SN}}
 \Delta\mu_i\big(C_{\mathrm{SN}}^{-1}\big)_{ij} \Delta\mu_j\ ,\nn
 \Delta\mu_i&=&\mu^\mathrm{th}(z_i,\theta_1,\dots)-\mu^\mathrm{obs}_i\ .
 \label{chiSN}\eea
 Here $\theta_j$ are free model parameters, $C_{\mbox{\scriptsize SN}}$ is the
 $N_{\mathrm{SN}}\times N_{\mathrm{SN}}$ covariance matrix and $\mu^\mathrm{th}$ are the theoretical values for the distance moduli, which is calculated as
follows:
\begin{equation}
 \mu^\mathrm{th}(z) = 5 \log_{10} \frac{(1+z)\,D_M(z)}{10\mbox{pc}},\qquad D_M(z)= c \int\limits_0^z\frac{d\tilde z}{H(\tilde
 z)}.    \label{muDM}
\end{equation}
For evaluating the function (\ref{chiSN}), the Hubble constant $H_0$ (or equivalently the ``asymptotical'' constant $H_0^*$) is considered  as a nuisance parameter.\\

For BAO new data from Dark Energy Spectroscopic Instrument (DESI) Data
Release 1 \cite{DESI:2024} is considered. The comparison is performed by calculating two distances:
\begin{equation}
d_z(z)= \frac{r_s(z_d)}{D_V(z)}\, ,\qquad A(z) = \frac{H_0\sqrt{\Omega_m^0}}{cz}D_V(z)\,
, \label{dzAz}
\end{equation}
where $D_V(z)=\big[{cz D_M^2(z)}/{H(z)} \big]^{1/3}$, $z_d$ being the redshift at the
end of the baryon drag era whereas the comoving sound horizon $r_s(z)$ is obtained as follows:
\cite{OdintsovSGS_Axi:2023}:
  \bea
r_s(z)&=&  \int_z^{\infty} \frac{c_s(\tilde z)}{H (\tilde z)}\,d\tilde z=\nn
&=&\frac1{\sqrt{3}}\int_0^{1/(1+z)}\frac{da}{a^2H(a)\sqrt{1+\big[3\Omega_b^0/(4\Omega_\gamma^0)\big]a}}\ .
  \label{rs2}\eea
The estimations for $z_d$ and for the ratio of baryons and photons $\Omega_b^0/\Omega_\gamma$ are fixed by the Planck 2018 data \cite{Planck18}.

DESI DR1 data \cite{DESI:2024} that includes BAO data from clustering of galaxies, quasars and the Lyman-$\alpha$ forest in the redshift range $0.1<z<4.16$  provides 6 datapoints, shown in Table~\ref{DESI}. In addition, another 21 BAO data points that provide $d_z(z)$ as well as 7 data points that gives $A(z)$ are considered \cite{OdintsovSGS_Axi:2023,OdintsovSGS_LnAx:2024,OdintsovOS:2023}. Then, the following $\chi^2$ function is obtained for the fittings with BAO data:
\begin{equation}
\chi^2_{\mathrm{BAO}}(\Omega_m^0,\theta_1,\dots)=\Delta d\cdot C_d^{-1}(\Delta d)^T +
\Delta { A}\cdot C_A^{-1}(\Delta { A})^T\, . \label{chiBAO}
\end{equation}
Here, $\Delta d_i=d_z^\mathrm{obs}(z_i)-d_z^\mathrm{th}(z_i,\dots)$, $\Delta
A_i=A^\mathrm{obs}(z_i)-A^\mathrm{th}(z_i,\dots)$, $C_{d}$ and $C_{A}$ are the
covariance matrices for the correlated BAO data \cite{Percival:2009,Blake:2011}.\\

\begin{widetext}
\begin{table}[hb]
\begin{tabular}{|c|c|c|c|c|c|c|}
\hline  $z_\mathrm{eff}$ & 0.295& 0.51 & 0.706 & 0.93 & 1.317 & 2.33\\
\hline
 $z$ range &0.1 - 0.4&0.4 - 0.6&0.5 - 0.8&0.8 - 1.1&1.1 - 1.6 &1.77 - 4.16\\
\hline
 $d_z$ &$0.1261 \pm0.0024  $&$0.0796 \pm0.0018  $ &$0.0629 \pm0.0014  $ &$0.05034\pm0.0008  $
&$0.04144\pm0.0011  $ &$0.03173\pm 0.00073$\\
\hline
 \end{tabular}
 \caption{DESI DR1 BAO data.}
\label{DESI}
\end{table}
\end{widetext}
Moreover, Cosmic Chronometers (CC) or the Hubble parameter data $H(z)$, which are measured as $H (z)=
\frac{\dot{a}}{a} \simeq -\frac{1}{1+z} \frac{\Delta z}{\Delta t}$ from differential
ages $\Delta t$ of galaxies with known $\Delta z$  are considered, which provides $N_H=32$ CC datapoints \cite{HzData}. The $\chi^2$ function for CC $H(z)$
data is:
\begin{equation}
    \chi_H^2(\theta_1,\dots)=\sum_{j=1}^{N_H}\bigg[\frac{H(z_j,\theta_1,\dots)-H^{obs}(z_j)}{\sigma _j}  \bigg]^2\ .
    \label{chiH}
\end{equation}

Regarding the CMB  observational parameters from Planck 2018 data \cite{Planck18}, the following values are considered \cite{ChenHuangW2018}:
  \begin{equation}
  R=\sqrt{\Omega_m^0}\frac{H_0D_M(z_*)}c,\quad
 \ell_A=\frac{\pi D_M(z_*)}{r_s(z_*)},\quad\omega_b=\Omega_b^0h^2\ ,
 \label{CMB} \end{equation}
where $z_*$ is the photon-decoupling redshift estimated in
Ref.~\cite{ChenHuangW2018}, $D_M$ is the comoving distance (\ref{muDM}),
$h=H_0/[100\,\mbox{km}\mbox{s}^{-1}\mbox{Mpc}^{-1}]$ and $r_s(z)$ is the comoving sound
horizon (\ref{rs2}).  The corresponding $\chi^2$ function for the CMB data is computed by:
 \begin{equation}
\chi^2_{\mbox{\scriptsize CMB}}=\min_{\omega_b}\Delta\mathbf{x}\cdot
C_{\mathrm{CMB}}^{-1}\big(\Delta\mathbf{x}\big)^{T},\qquad \Delta
\mathbf{x}=\mathbf{x}-\mathbf{x}^{Pl}
 \label{chiCMB} \end{equation}
where $\mathbf{x}=\big(R,\ell_A,\omega_b\big)$. The observational values, which are obtained with free amplitude for the lensing power spectrum, are provided in \cite{Planck18,ChenHuangW2018}.
 \be
 \mathbf{x}^{Pl}=\big(1.7428\pm0.0053,\;301.406\pm0.090,\;0.02259\pm0.00017\big)\ .
\ee
The covariance matrix $C_{\mathrm{CMB}}=\|\tilde C_{ij}\sigma_i\sigma_j\|$ is described in Ref.~\cite{ChenHuangW2018}.\\

In the following section, the model (\ref{FR2}) and the $\Lambda$CDM model are compared by using the above observational data sources and the corresponding fits for the free parameters are obtained.

\section{Testing the models with observational data}
\label{Results}

The $F(R)$  model (\ref{FR2}) is compared to the above observational data. To do so, two cases are considered: fixing $\alpha=1$ and keeping $\alpha$ as a free parameter  its observational predictions. For this purpose, the total $\chi^2$ function with the contributions from SN Ia, BAO, CC and CMB is computed:
 \begin{equation}
  \chi^2=\chi^2_\mathrm{SN}+\chi^2_\mathrm{BAO}+\chi^2_H+\chi^2_\mathrm{CMB}\ .
 \label{chitot} \end{equation}

The model (\ref{FR2}), after fixing the radiation-matter ratio (\ref{Xrm}), contains $N_p=5$ free parameters:
 \begin{equation}
 \alpha, \quad\beta, \quad \Omega_m^0, \quad \Omega_\Lambda,\quad H_0\,.
\label{FreeParam} \end{equation}
For $\alpha=1$, that recovers the standard exponential gravity model, the number of free parameters is reduced to $N_p=4$. In both scenarios the fittings are computed for the parameters  $\Omega_m^0$, $\Omega_\Lambda$, $H_0$ instead of $\Omega_m^*$,
$\Omega_\Lambda^*$, $H_0^*$, by using the relations (\ref{H0Omm}).

\begin{widetext}

 \begin{figure}[th]
   \centerline{ \includegraphics[scale=0.66,trim=5mm 0mm 2mm -1mm]{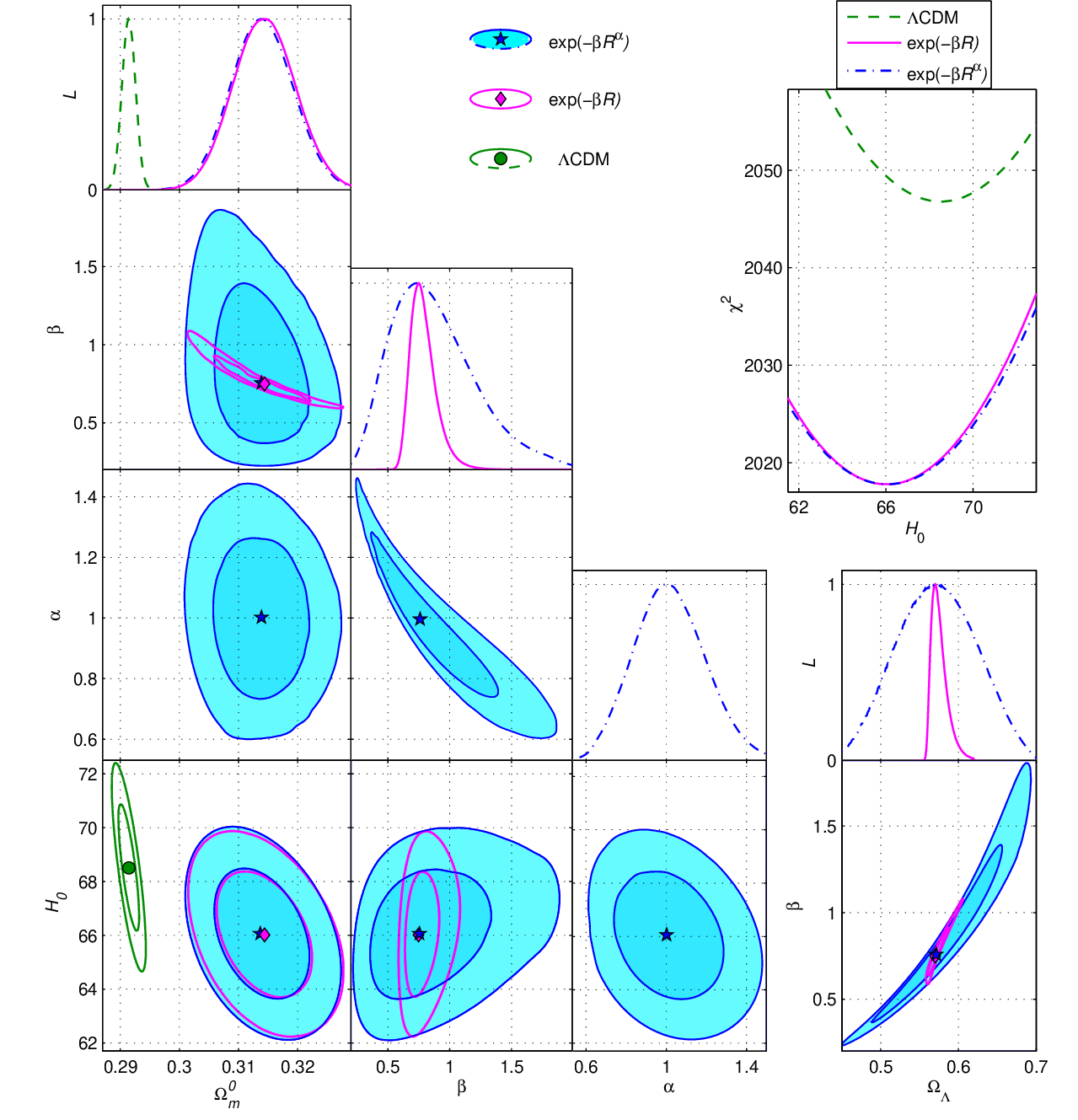}}
\caption{Contour plots of $\chi^2$ with $1\sigma$, $2\sigma$ CL, likelihood functions $
{\cal L}(\theta_i)$ and one-parameter distributions $\chi^2(H_0)$ for the  exponential
$F(R)$ model  (\ref{FR2}) in comparison with its particular case $\alpha=1$ and with the
$\Lambda$CDM model. }
  \label{F1}
\end{figure}
\end{widetext}

The results for both  exponential cases are depicted in Fig.~\ref{F1}, where the
$\Lambda$CDM model is also included for comparison, which is described by the Hubble
parameter:
 \be H^2=H_0^2\big[\Omega_m^{0}(a^{-3}+ X_r
a^{-4})+1-\Omega_m^0-\Omega_r^0\big]\ ,
 \label{HLCDM} \ee
which contains $N_p=2$ free parameters $\Omega_m^0$  and $H_0$. Note that the
$\Lambda$CDM scenario is recovered in the limit $\beta\to+\infty$ for the model
(\ref{FR2}) independently of $\alpha>0$.

In order to show the results of the fits for the free parameters (\ref{FreeParam}),  the corresponding
 contour plots are depicted in Fig.~\ref{F1}. The contours correspond to $1\sigma$
(68.27\%) and $2\sigma$ (95.45\%) confidence regions for the two-parameter distributions
$\chi^2(\theta_i,\theta_j)$, which are obtained by minimising the $\chi^2$ over all the
remaining free parameters. For instance, the contours depicted in the bottom-left panel of Fig.~\ref{F1} are obtained by
 $$\chi^2(\Omega_m^0,H_0)=\min\limits_{\alpha,\beta,\Omega_\Lambda}\chi^2(\alpha,\dots,H_0)\;,$$
  and show the estimations for the three models.

The blue stars for the $F(R)$ model (\ref{FR2}), the magenta diamonds for the case
$\alpha=1$ and the green circles for $\Lambda$CDM denote the best fits with $\min\chi^2$
of the corresponding two-dimensional distributions. The best fits for the free
parameters are also shown in Fig.~\ref{F1} in the one-parameter distributions
$\chi^2(H_0)$  and the likelihoods ${\cal L}(\theta_j)$, which are obtained by:
   \begin{equation}
{\cal L}(\theta_j)= \exp\bigg[- \frac{\chi^2(\theta_j)-m^\mathrm{abs}}2\bigg]\ ,
 \label{likeli} \end{equation}
where $\chi^2(\theta_j)=\min\limits_{\mbox{\scriptsize other
}\theta_k}\chi^2(\theta_1,\dots)$, $\theta_j$ is the corresponding model parameter and
$m^\mathrm{abs}$ the absolute minimum for $\chi^2$. In addition, the best fits with the
corresponding $1\sigma$ errors for the free model parameters (\ref{FreeParam})  are also
shown explicitly in Table~\ref{Estim} below.

As shown in Table~\ref{Estim}, the generalized exponential $F(R)$  model (\ref{FR2})
with provides a best fit for $\alpha$ that is very close to 1, in other words, for the
above observational data, the generalized  model (\ref{FR2}) achieves the best results
when it becomes the standard exponential $F(R)$  model (\ref{FR2}) with $\alpha=1$. In
addition, both cases show the same absolute value for the   $\min\chi^2$.

The one-parameter distributions $\chi^2(H_0)$ for all the models are shown in the
top-right panel of Fig.~\ref{F1}. The results seem to favour clearly the modified
gravity models in comparison to the  $\Lambda$CDM model. One may conclude that the last
Pantheon+ SN Ia and DESI BAO observational data change the domain of the free
parameters, where both exponential $F(R)$ models achieve the most successful results, in
comparison to previous analysis in the literature \cite{OdintsovSGS:2017}. The fits for
the parameter $\beta$ favours smaller values than previous analysis with older data. The
main point lies on the fact that the smaller $\beta$ values are, the larger the
difference  between exponential $F(R)$ models and the $\Lambda$CDM arise, which is
reflected in the different values for $\chi^2$ for each model. By considering larger
$\beta$ values, the $\chi^2$ in the $F(R)$ cases become larger and tend to the
$\Lambda$CDM results.

In addition, the different beahviour of the exponential $F(R)$ models in comparison to
the $\Lambda$CDM scenario also leads to different predictions for the best fits of the
Hubble constant $H_0$ and for the matter density parameter $\Omega_m^0$.
Table~\ref{Estim} shows that the  $\Lambda$CDM best fit of the Hubble constant is given
by $H_0=68.51^{+1.56}_{-1.53}$ km/(s$\cdot$Mpc) whereas for the model (\ref{FR2}), it
leads to $H_0=66.06^{+1.61}_{-1.59}$ km/(s$\cdot$Mpc) (the case $\alpha=1$ provides
similar result), which are mutually excluded at $1\sigma$.
 For the matter density parameter $\Omega_m^0$ the
$\Lambda$CDM  and $F(R)$ models are excluded to more than  $3\sigma$ in their
predictions, as shown in Fig.~\ref{F1}.

\bigskip

The best fits for $\beta$  and  $\Omega_\Lambda$ are rather close for both exponential
models, in particular,  $\Omega_\Lambda=0.571^{+0.058}_{-0.057}$ for model (\ref{FR2})
and $\Omega_\Lambda=0.570^{+0.010}_{-0.007}$ for its case $\alpha=1$. For the scenario
(\ref{FR2}), the $1\sigma$ error are larger (and the contours are wider) that may be
connected with the additional degree of freedom $\alpha$, though the best fit for
$\alpha$ is close to 1. However, they both lead to similar best fits for $H_0$ and
$\Omega_m^0$.

The large difference regarding the best fits for both exponential models when comparing the absolute minimum $m^\mathrm{abs}=\min\chi^2$
with respect to the standard $\Lambda$CDM model does not vanish even when considering the number of free parameters $N_p$ for each case and following the Akaike
information criterion \cite{Akaike}
 \be
 \mbox{AIC} = \min\chi^2_{tot} +2N_p.
  \label{AIC}\ee

The more free parameters has a model, the larger AIC is. As shown in Table \ref{Estim},
despite $2N_p$ is smaller for the $\Lambda$CDM model, this does not save the standard
model of cosmology in comparison to the exponential modified gravity. To show this clearer, the difference
$\Delta\mbox{AIC}=\mbox{AIC}_\mathrm{model}-\mbox{AIC}_{\Lambda\mathrm{CDM}}$ is included in Table \ref{Estim}.

\section{Parameterisations of the Dark Energy EoS}
\label{CPL}

Previous results concerning the great success of the exponential $F(R)$ models in comparison to
$\Lambda$CDM model when considering the Akaike information criterion suggests two key questions: (a)
what is the nature of the large differences according to $\Delta\mbox{AIC}$, are any of the
observational data strongly favouring modified gravity in comparison to others?; (b) are those results pointing to a dynamical EoS parameter for dark energy? The last question is also inspired by recent attempts
\cite{DESI:2024,Chudaykin:2024,Park:2024brx,Notari:2024} to confront the last DESI BAO
data with different scenarios.

In this section, we test two popular parameterisations of EoS for
dark energy: the so-called $w$CDM model and the Chevallier-Polarski-Linder (CPL or $w_0w_a$CDM) model
\cite{CPL}. The pressure and energy densities of dark energy are related by its EoS
$$ p_\mathrm{de}=w \rho_\mathrm{de}\;,$$
where,
\bea 
w&=&\text{const}\ , \quad \text{for}\quad w-\text{CDM model}\ ,\nn
w&=&w_0+w_1(1-a)\ , \quad \text{for CPL model}\ . 
\eea
Both provide a generalisation of $\Lambda$CDM model in an agnostic model way. Then, the
Hubble parameter yields:
 \be H^2=H_0^2\big[\Omega_m^{0}(a^{-3}+ X_r
a^{-4})+(1-\Omega_m^0-\Omega_r^0)\,f(a)\big]\ ,
 \label{HCPL} \ee
where $f(a)=a^{-3(1+w)}$ for $w$CDM and $f(a)=a^{-3(1+w_0+ w_1)}e^{3w_1(a-1)}$ for CPL
model. By comparing to $\Lambda$CDM model, one can see that an additional parameter $w$ arises in
$w$CDM while CPL model contains two extra free parameters, $w_0$, $w_1$. For $w_1=0$ CPL reduces to
$w$CDM whereas $\Lambda$CDM model is recovered for $w=w_0=-1$  in both cases (together with $w_1=0$ in the CPL model).

\begin{widetext}

 \begin{figure}[th]
   \centerline{ \includegraphics[scale=0.66,trim=5mm 0mm 2mm -1mm]{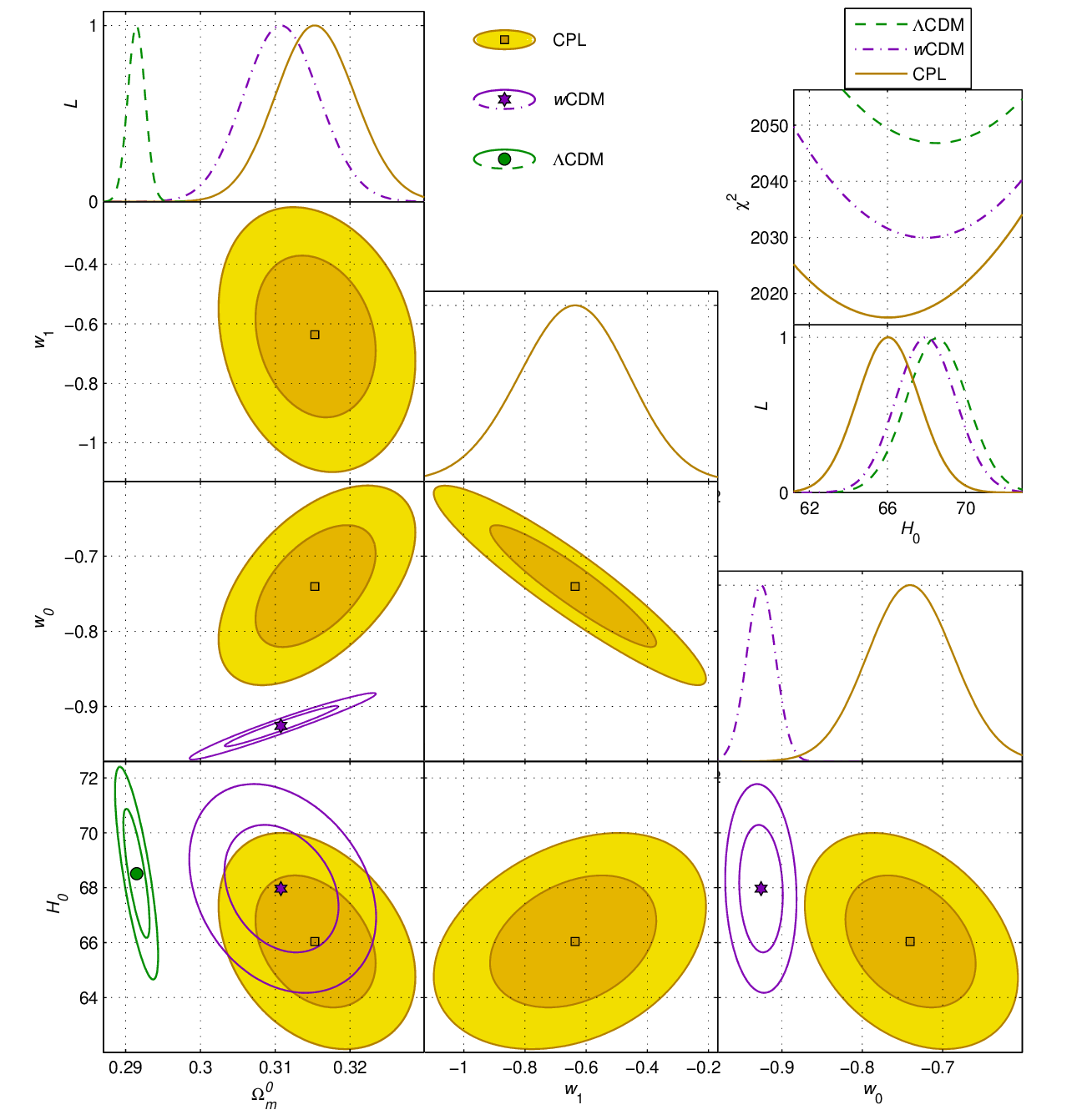}}
\caption{Contour plots of $\chi^2$ with $1\sigma$, $2\sigma$ CL, likelihoods $ {\cal
L}(\theta_i)$ and one-parameter distributions $\chi^2(H_0)$ for the $w$CDM and CPL
scenarios in comparison with the $\Lambda$CDM model. }
  \label{F2}
\end{figure}
\end{widetext}

Results of the comparison with observational data for these models are shown in Table \ref{Estim} and depicted
in Fig.~\ref{F2}. For  $w$CDM  model the absolute minimum
of $\chi^2$, the value for AIC (\ref{AIC}) and the best fits for $\Omega_m^0$ and $H_0$ lies in-between
the  $\Lambda$CDM and $F(R)$ results, whereas data again strongly favours the CPL  scenario, where the $\min\chi^2$ and AIC parameter achieve the smallest values.

\begin{widetext}

\begin{table}[ht]
\begin{tabular}{|l|c|c|c|c|c|c|}
\hline  Model &   $\min\chi^2/d.o.f$& AIC & $\Delta$AIC& $\Omega_m^0$& $H_0$& other parameters  \\
\hline
 (\ref{FR2}):$\,e^{-\beta{\cal R}^\alpha}$ & 2017.80 /1766 & 2027.80&$-22.99$&$0.3138^{+0.0054}_{-0.0052}$
& $66.06^{+1.61}_{-1.59}$ & $\beta=0.733^{+0.377}_{-0.273}$, $\alpha=1.002^{+0.184}_{-0.173}$  \rule{0pt}{1.1em}  \\
\hline
 Exp  $e^{-\beta{\cal R}}$& 2017.80 /1767 & 2025.80&$-24.99$& $0.3144^{+0.0053}_{-0.0055}$ & $66.02^{+1.57}_{-1.53}$& $\beta=0.750^{+0.099}_{-0.079}$ \rule{0pt}{1.1em}  \\
\hline
$\Lambda$CDM& 2046.79 /1769 & 2050.79& 0 & $0.2914^{+0.0012}_{-0.0011}$& $68.51^{+1.56}_{-1.53}$& - \rule{0pt}{1.1em}  \\
\hline
 $w$CDM& 2029.93 /1768 & 2035.93&$-14.86$& $0.3108^{+0.0050}_{-0.0050}$& $67.97^{+1.53}_{-1.53}$\rule{0pt}{1.1em} & $w=-0.926^{+0.018}_{-0.018}$   \\
\hline
 CPL& 2015.72 /1767 & 2023.72 &$-27.07$&$0.3153^{+0.0053}_{-0.0053}$&$66.03^{+1.59}_{-1.58}$\rule{0pt}{1.1em} &$\;w_0=-0.741^{+0.054}_{-0.053},\;\;
w_1=-0.635^{+0.175}_{-0.183}$ \\
\hline
 \end{tabular}
 \caption{The best fit values for the free parameters,  $\min\chi^2$, AIC and $\Delta$AIC for the two exponential $F(R)$ models (\ref{FR2})
 in comparison with $\Lambda$CDM, $w$CDM and CPL models.}
\label{Estim}
\end{table}
\end{widetext}

As shown in Table \ref{Estim}, the best fits of $\Omega_m^0$ and $H_0$ for the CPL model and exponential
$F(R)$ gravities are very close, such that the CPL best fits also differs with respect to the
$\Lambda$CDM predictions for more than $3\sigma$ in the case of the density parameter $\Omega_m^0$ and near $1\sigma$ for
$H_0$.

According to $\min\chi^2$ and the Akaike information criterion, CPL fits slightly better the data in comparison to the standard exponential gravity model, but with no statistical significance, despite both models contain the
same number of free parameters $N_p=4$. However, this difference in AIC and
$\Delta\mbox{AIC}$ is small in comparison with that for $w$CDM and $\Lambda$CDM. The
generalised exponential $F(R)$ model (\ref{FR2}) shows a worse AIC comparison with respect to standard exponential gravity and CPL model, since it contains an extra free parameter
($N_p=5$).

The CPL best fits and $\min\chi^2$ values in Table \ref{Estim} and Fig.~\ref{F2} differ
from the $w$CDM results, because the CPL best fit for $w_1$ is far from zero, which corresponds to $w$CDM
model. Indeed, the smaller AIC value for the CPL model reflects a strong deviation from a constant EoS for dark energy, either $w$CDM or $\Lambda$CDM model. Moreover, $w$CDM lies in-between the CPL model and $\Lambda$CDM model concerning the goodness of the fits as provided by the AIC parameter. Also the value for the Hubble constant is close for both cases.
\\

\begin{figure}[th]
   \centerline{ \includegraphics[scale=0.65,trim=-4mm 0mm 3mm -1mm]{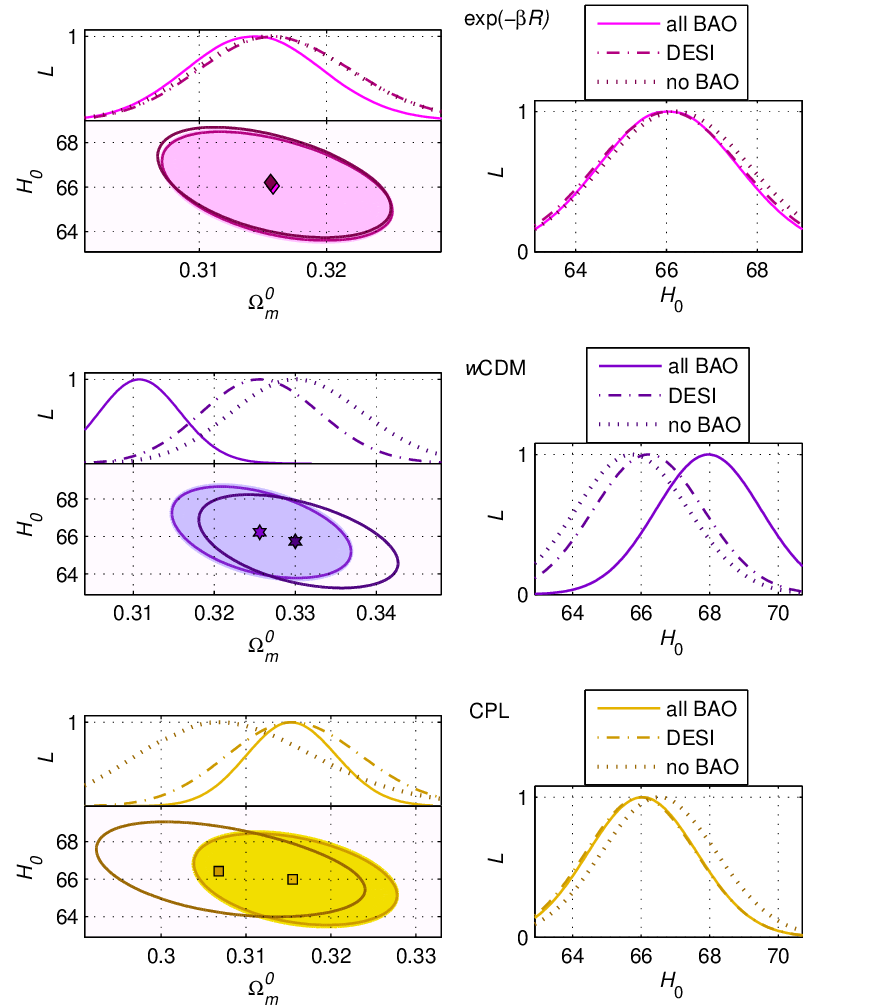}}
\caption{Comparison of $1\sigma$ contour plots and likelihoods  for the exponential
$F(R)$, $w$CDM, $\Lambda$CDM and CPL scenarios, filled and empty contours correspond to
only DESI and without BAO data respectively. }
  \label{F3}
\end{figure}

The question about the origin of the large devitations $\Delta\mbox{AIC}$ favouring the standard exponential gravity
and the CPL model requires additional investigations: what type of new observational data
is behind these results? It may be connected with the DESI DR1 BAO data
\cite{DESI:2024} that are tabulated as 6 datapoints in Table~\ref{DESI}. To understand their
role, additional tests are followed for the standard exponential  $F(R)$ model, $w$CDM,
CPL and $\Lambda$CDM scenarios by considering two sets of observational data separately: (a) SN Ia, $H(z)$
or Cosmic Chronometers (CC), CMB  and only 6 DESI BAO;
and (b) SN Ia, CC, CMB observational data (without BAO). The results are
shown in Table \ref{Estim2} and Fig.~\ref{F3}. The generalised exponential model (\ref{FR2}) is excluded because of the large number of free parameters
$N_p=5$ that increases the AIC value in comparison to standard exponential gravity.

\begin{widetext}

\begin{table}[ht]
\begin{tabular}{|l|c|c|c|c|c||c|c|c|c|c|}
\hline
 Data & \multicolumn{5}{c||}{(a) \ SN Ia + CC + CMB + 6 DESI BAO datapoints}&   \multicolumn{5}{c|}{(b) \ SN Ia + CC + CMB \rule{0pt}{1.1em} }\\
\hline
Model &   $\min\chi^2/d.o.f$& AIC & $\Delta$AIC& $\Omega_m^0$& $H_0$&  $\min\chi^2$& AIC & $\Delta$AIC& $\Omega_m^0$& $H_0$  \\
\hline
Exp $e^{-\beta{\cal R}}$& 2000.32 /1746 & 2008.32&$-27.83$& $0.3158^{+0.0061}_{-0.0058}$ & $66.05^{+1.58}_{-1.63}$& 1997.98 & 2005.98&$-28.24$& $0.3156^{+0.0061}_{-0.0060}$ & $66.21^{+1.63}_{-1.63}$ \rule{0pt}{1.1em}  \\
\hline
$\Lambda$CDM& 2032.15 /1748 & 2036.15& 0 & $0.2913^{+0.0012}_{-0.0012}$& $68.60^{+1.62}_{-1.59}$& 2030.22 & 2034.22& 0 & $0.2912^{+0.0012}_{-0.0012}$& $68.72^{+1.64}_{-1.60}$ \rule{0pt}{1.1em}  \\
\hline
 $w$CDM& 2005.44 /1747 & 2011.44&$-24.71$& $0.3256^{+0.0074}_{-0.0072}$& $66.22^{+1.61}_{-1.61}$\rule{0pt}{1.1em} &2001.59 & 2007.59&$-26.63$& $0.3300^{+0.0083}_{-0.0079}$& $65.74^{+1.64}_{-1.60}$  \\
\hline
 CPL& 1998.82 /1746 & 2006.82 &$-29.33$&$0.3155^{+0.0080}_{-0.0077}$&$65.99^{+1.61}_{-1.61}$\rule{0pt}{1.1em} &1995.68 & 2003.68&$-30.54$& $0.3068^{+0.0110}_{-0.0101}$&
 $66.43^{+1.71}_{-1.63}$
 \\
\hline
 \end{tabular}
 \caption{The best fits,  $\min\chi^2$, AIC and $\Delta$AIC for observational data (a) with only 6 DESI BAO
datapoints and (b) without BAO data for the exponential $F(R)$ gravity, $\Lambda$CDM, $w$CDM and
CPL models.}
\label{Estim2}
\end{table}
\end{widetext}

By analyzing the results shown in Table \ref{Estim2}, one may conclude that the DESI BAO data can slightly
shift some estimations of the model parameters, but does not affect the large
difference $\Delta$AIC between $\Lambda$CDM by one side and exponential, $w$CDM or CPL
models by the another side. Indeed, such a difference becomes larger in absence of DESI BAO data We see that  $\Delta$AIC  between the mentioned 3 scenarios do
not disappear in the cases (a) and (b). Thus, the reason of large differences in $\Delta$AIC of the exponential, $w$CDM and
CPL scenarios lies in the SN Ia Pantheon+  data \cite{PantheonP:2022}. Remind that a large
negative $\Delta$AIC of the aforementioned 3 models points to a clear advantage in describing the observational data in comparison with $\Lambda$CDM.

By following the hierarchy of AIC criterion for these four models when considering the three observational datasets (with all BAO data, just with DESI BAO data and with no BAO data), shown in Tables~\ref{Estim} and \ref{Estim2}, CPL model arises as the most successful one, followed by the exponential
$F(R)$ model, $w$CDM and $\Lambda$CDM. However, these models respond differently to changes in
BAO data sets. For the exponential and $\Lambda$CDM models, $\min\chi^2$ and AIC grow
successively when BAO datasets are considered, while the best fits of the density parameter $\Omega_m^0$ and the Hubble constant $H_0$ remain nearly the same for all three datasets, as shown in Fig.~\ref{F3}.

Another picture takes place for $w$CDM and CPL scenarios. On can see fundamental changes
of the best fits for $\Omega_m^0$ and $H_0$  in Fig.~\ref{F3}, Tables~\ref{Estim} and \ref{Estim2}, where $w$CDM model shows a remarkable difference on the $\min\chi^2$ and AIC when DESI data is included and especially for the other BAO data. The corresponding changes of  AIC for
$w$CDM from $-26.63$ (without BAO data) to  $-14.86$ (with BAO data) are essentially
larger than for the other models. One also can see that the best fits for
$\Omega_m^0$ for the  $w$CDM  model leads to important differences when BAO data is included (while the best fit for $H_0$ grow). The parameters distributions for the CPL model also differs from one dataset to another but with smaller deviations, whereas exponential gravity points to the same fits independently of the BAO datasets.

%
\section{Conclusions}
\label{conclusions}
%

As shown in previous literature, exponential $F(R)$ gravities have been widely studied,
since they show an ability to reproduce well the whole cosmological evolution, including
GR predictions at both early as late times, and at the same time they are constructed in
such a way that pass local scale constraints  \cite{Linder2009,OdintsovSGS:2017}.

In this paper, exponential $F(R)$ gravity is reconsidered and confronted to the latest
observational data, including  Pantheon+ SN Ia, new BAO data from DESI DR1, the Hubble
parameter CC estimations and CMB data. To do a reliable test with the aforementioned
observational data, a generalised exponential gravity model is also considered to check
deviations from standard exponential gravity. Both models are then compared with the
fits and predictions of the standard $\Lambda$CDM model and also to $w$CDM and CPL models \cite{CPL}.

 To do the fits, the usual technique of the minimum $\chi^2$ is followed. The best fits
for these scenarios are summarised in Table \ref{Estim} and also shown in
Figs.~\ref{F1}, \ref{F2}. As one can see, both exponential gravity models show a much
better accuracy while fitting the observational data than $\Lambda$CDM model, as
indicated by a much smaller $\min\chi^2$. Moreover, even if one considers the extra free
parameters of the modified gravity models as an artificial handmade way to get better
fits, the observational data still favours clearly these models to the detriment of
$\Lambda$CDM model. Indeed, when computing the AIC coefficient which penalise the number
of free parameters, results still support strongly the modified gravity scenario. In
addition, the fits of the generalised exponential gravity model points towards the
standard exponential gravity case,
since the extra parameter is centred in $\alpha=1$, which recovers the usual exponential gravity model. In addition, the standard exponential gravity model is also strongly
supported in comparison with  $w$CDM model, but slightly below the CPL model with the observational data.\\

In fact, previous analysis of the same exponential gravity model favoured $\Lambda$CDM,
as shown in Ref.~\cite{OdintsovSGS:2017}, where the results led to a value for the
fundamental parameter given by $\beta=2.38^{+\infty}_{-0.8}$. Recalling that
$\Lambda$CDM is recovered for $\beta\rightarrow\infty$, those previous results were
clearly pointing out to $\Lambda$CDM model, which was included within $1\sigma$
confidence region for $\beta$. Here, we have shown that this is not the case for the
last available observational data, where one obtains $\beta=0.75^{+0.099}_{-0.079}$ for
the standard exponential gravity model ($\alpha=1$) and  $\beta=0.733^{+0.377}_{-0.273}$
for the generalised model, both given by (\ref{FR2}). For the former, $\Lambda$CDM turns out excluded at $4\sigma$. In addition, in order to get a strong confidence
on these results, we have computed the $\min\chi^2$ by fixing the $\beta$ parameter to
larger and larger values, which at the end tend to the $\min\chi^2$ of $\Lambda$CDM
results.  From Table \ref{Estim} and  Fig.~\ref{F1}, one can also infer that the best
fits for the parameters $H_0$ and $\Omega_m^0$ in the exponential $F(R)$ models and in
the $\Lambda$CDM scenario differ notably. In particular, our results point to an even
smaller value for the Hubble constant $H_0=66.06^{+1.61}_{-1.59}$ km/(s$\cdot$Mpc), what
would increase the $H_0$ tension when compared with the value provided by the
calibration of SNe Ia \cite{SH0ES:2024,Riess:2021jrx,Brout:2022vxf}. \\

In addition, CPL parametrisation of the EoS parameter for dark energy also points to exclude $\Lambda$CDM at $4\sigma$, while the best fit for $H_0$ in this scenario is also small and almost coincides with the
 $F(R)$ prediction. Nevertheless, the Hubble constant for the exponential gravity and CPL fits better the
one provided by Planck estimations from the CMB  \cite{Planck18}. For the matter density
parameter $\Omega_m^0$, a difference of more than $3\sigma$ arises in comparison of the
$\Lambda$CDM  with the $F(R)$ models and also with CPL, what results in the condition of mutually
exclusive of both descriptions of the cosmological evolution. Moreover, the goodness of the fits favours clearly a dynamical EoS for dark energy, either is described by a parameterisation (CPL model) or by modified gravity.\\

Hence, tests with the latest observational data show a large advantage of the exponential
$F(R)$ model,  CPL and $w$CDM scenarios in comparison to $\Lambda$CDM model in terms of the minimum
of $\chi^2$ and the AIC criterion. This picture strongly differs from previous results with older observational data \cite{OdintsovSGS:2017, OdintsovSGStens:2021,OdintsovOS:2017log,OdintsovSGSlog:2019}. To find out the answer behind this change of paradigm, additional tests are raised for every model, where by assuming the observational data provided by SN Ia, CC and CMB, we played with the different sets of BAO data: (a) with
only 6 DESI BAO datapoints and (b) without BAO data. By Analysing the
results shown in Table \ref{Estim2} and Fig.~\ref{F3}, we may conclude that the the large difference in the AIC parameter between $F(R)$, CPL, $w$CDM models and
$\Lambda$CDM scenarioa are not connected not with DESI BAO data, but with SN Ia Pantheon+  data \cite{PantheonP:2022}.\\

\


Then, success of the exponential gravity in describing the observational datasets suggests that a more complete theory of gravity, beyond GR, have to include non-trivial
terms of the Ricci scalar in the action at the cosmological limit, at least. Hence, we can conclude that the game of rivalry among modified gravities/dynamical dark energy vs $\Lambda$CDM is over now.

\section*{Acknowledgments}

This work was  supported by MICINN (Spain) project PID2020-117301GA-I00 (D.S.G.) funded by MCIN/AEI/10.13039/501100011033 (``ERDF A way of making Europe" and ``PGC Generaci\'on de Conocimiento") and also by the program Unidad de Excelencia Maria de Maeztu CEX2020-001058-M, Spain (S.D.O).

\end{document}